\documentclass[a4paper]{jpconf}
\usepackage{graphicx}
\begin{document}
\title{A numerical study on the discharges in Micromegas}

\author{Deb Sankar Bhattacharya, Raimund Str{\"o}hmer, Thomas Trefzger}

\address{ University of W{\"u}rzburg, Physik und ihre Didaktik, Emil-Hilb-Weg 22, 97074, Germany}

\ead{deb.sankar.bhattacharya@gmail.com}

\begin{abstract}
The Micro-Pattern Gaseous Detectors (MPGD) have been widely adopted in nuclear and particle physics experiments, for their fast response and other excellent characteristics. To achieve the required signal strength and detection efficiency, sometimes they are operated at a high voltage range. This often challenges the limit of high voltage stability of the detector. Discharge in gaseous detectors is a complex process and involves several responsible factors. The microscopic geometrical structures of the MPGDs may itself sometimes induce discharges.

In this study, we are numerically investigating the discharge phenomena in non-resistive Micromegas. Within the COMSOL framework, a 3-dimensional model is developed to observe the occurrence and the development of discharge in Micromegas. The effect of space charge has been taken into account in the calculation. The model allows to vary the geometrical parameters of the detector as well as to study the effects of gas impurities and different number of primary charges. 

\end{abstract}

\section{Introduction}
Gaseous detector has always been an efficient choice where large area coverage, low material budget or continuous tracking are the primary concerns. After its invention, the gas amplification and the readout technology have been evolving during last century. Presently, with the advent of photolithography and very compact and fast electronics, the gaseous detectors have reached the era of Micro-Pattern Gaseous Detectors (MPGDs) [1]. Hybridization of different MPGDs is another trend of R$\&$D of present time. Fast response, high radiation hardness, robust structure, low ion feedback, very good spatial and time resolution, are salient excellent features of the MPGDs. With all these qualities, the MPGDs are becoming an important part of many big detectors like ATLAS [2], CMS [3], ALICE [4], ILD [5] and many others. 

\begin{figure}[h]
\begin{minipage}{12pc}
\includegraphics[width=11pc]{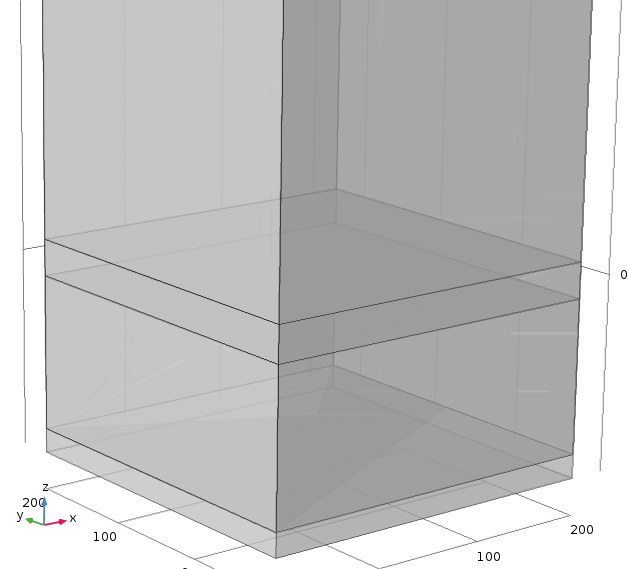}
\caption{\label{label}Parallel plates geometry. Cathode not shown.}
\end{minipage}\hspace{0.5pc}%
\begin{minipage}{12pc}
\includegraphics[width=12pc]{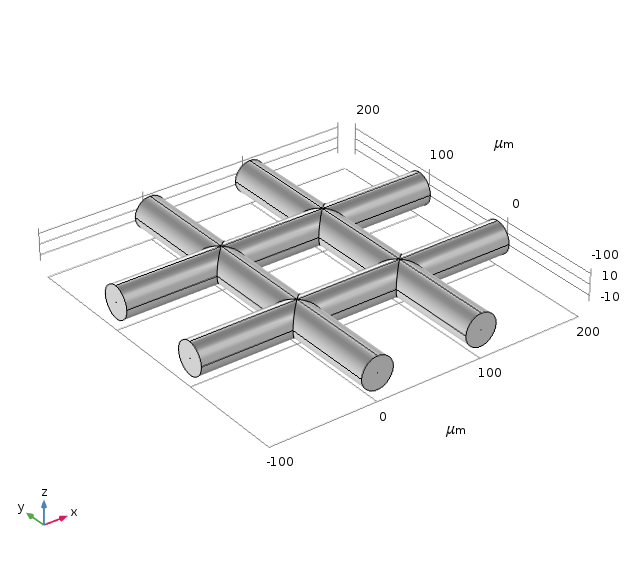}
\caption{\label{label}Calendared mesh geometry}
\end{minipage}\hspace{2pc}%
\begin{minipage}{12pc}
\includegraphics[width=11pc]{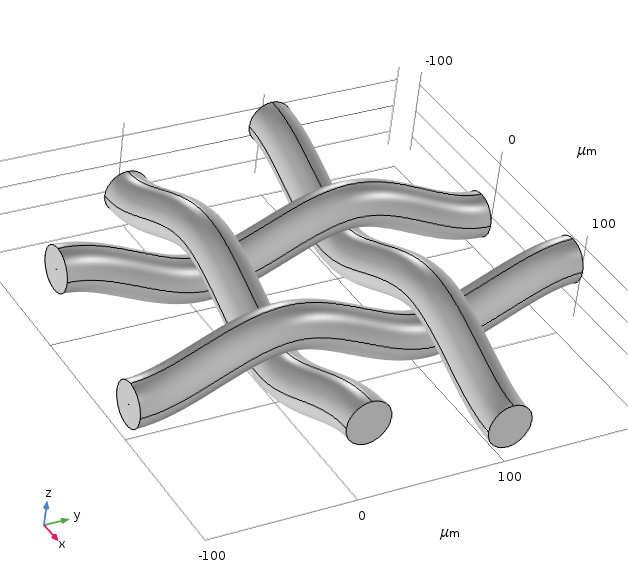}
\caption{\label{label}Woven mesh geometry}
\end{minipage}
\end{figure}

The charge amplification in a gaseous detector sometimes leads to uncontrolled avalanche or discharge, specially when they are operated close to their high voltage limit. Discharge can be seen in all kinds of gaseous detectors and there are several factors responsible for such a phenomena. Too frequent discharges or discharges with large amplitude can damage the detectors, moreover, it affects the measurement. Therefore, systematic studies on discharge in different MPGDs are truly relevant. Our present report is more focused on a particular type of MPGD, Micromegas [6]. We aim to study the electric field structure of different Micromegas geometry and also to build a 3-dimensional numerical model for discharge. The influence of the primary charge and gas impurities on the occurrence of discharge have been investigated by using a small gap (like in MPGD) parallel plate geometry. Moreover, the occurrence of discharge in two different Micromegas geometry has also been studied. 

\section{Modeling the problem and the simulation framework}
The simulation is conducted within the framework of COMSOL Multiphysics [7]. It essentially has two main parts: solving the electrostatic field and implementation of the field dependent charge transportation and charge multiplication. All the charge transport properties, as a function of field, are calculated with the Magboltz toolkit [8] and then incorporated in the framework. To solve the electric field for the detector geometry, the default field solver of COMSOL is used. It relies on Finite Elements Method. 
A simplified hydrodynamic model is considered to simulate discharge. The model follows a similar approach as was done by P. Fonte [9]. In a few recent studies [10], which apply the same method, the occurrence and propagation of discharge in 2-dimensional axis symmetric GEM structure [11] have been successfully illustrated. 

The following sets of equations have been solved in COMSOL to model discharge:

\begin{itemize}
\item electro statics:
\end{itemize}

\begin{equation}
\vec E = - \vec \nabla \phi
\end{equation}

\begin{equation}
\vec \nabla \cdot \epsilon \vec E = \rho_v
\end{equation}

\begin{itemize}
\item charge transport:
\end{itemize}

\begin{equation}
\frac{\partial C_i}{\partial t} + \vec \nabla \cdot (-D_i \vec \nabla C_i) + \vec u_i \cdot \vec \nabla C_i= R
\end{equation}

where,
\begin{equation}
R = n_e u (\alpha - AC - X) 
\end{equation}

Here, $\phi$ and $\vec E$ are respectively electric potential and electric field. $\rho_v$ is the volume density of charge. Equation 2 accounts for the space charge density developed in the detector. 

In equation 3, $C_i$ is the number density of charge, $D_i$ is the diffusion tensor, $u$ is the electron drift velocity, and $R$ is the rate of production of charge. $R$ is explained in equation 4, where, $n_e$ is the number density of electrons, $\alpha$ is the Townsend coefficient, $AC$ is the attachment coefficient, $X$ is the recombination rate. Here, the index $i$ is used to denote two different species, i.e., electrons and ions. Ions are slower than electrons and their mobility is taken to be constant here. The other transport parameters for the electrons are given as a function of electric field. Their dependence with electric field is computed in Magboltz and then plugged in to COMSOL. Equation 1-2 and equation 3 are inter-coupled. The charge moves according to the electric field and eventually multiplication starts. As the ions are slow, the space charge starts to grow and it changes the field. As the field changes the movement of charge changes accordingly. 

\begin{figure}[h]
\begin{minipage}{14pc}
\includegraphics[width=16pc]{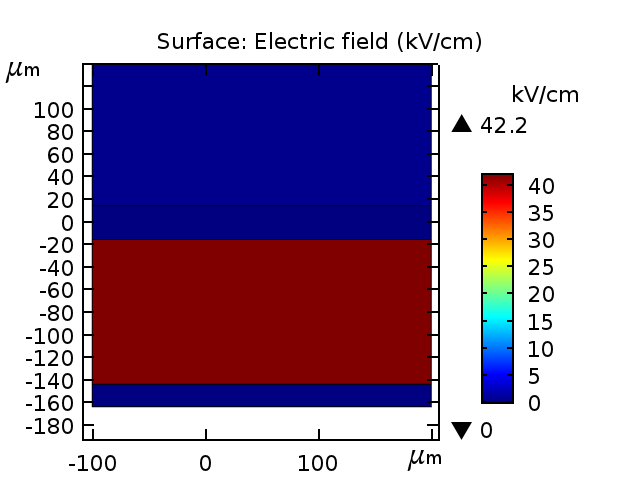}
\caption{\label{label} Surface plot of the field for the parallel plates.}
\end{minipage}\hspace{4pc}%
\begin{minipage}{14pc}
\includegraphics[width=16pc]{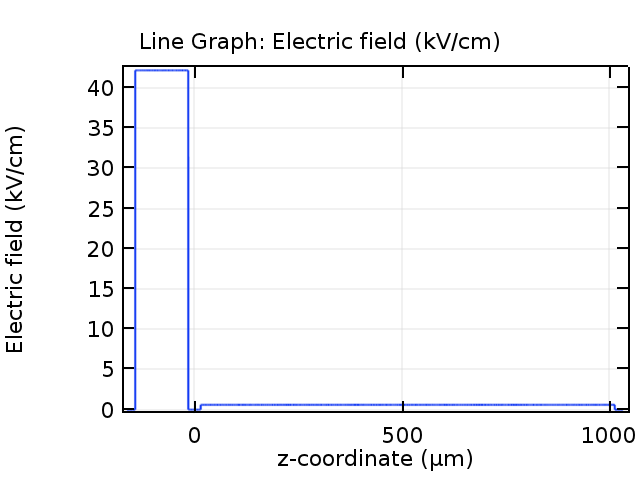}
\caption{\label{label} Line plot of the field for the parallel plates.}
\end{minipage} 
\end{figure}

\section{Simulation Configurations}
The geometry of a micro-mesh structure does not have a rotational symmetry (in $\phi$) and which is why just building a 2-D axis symmetric geometry does not necessarily correspond to a realistic 3-D Micromegas. Now, simulating the exact morphology of the detector geometry is a too ambitious goal. Therefore, for our calculations, full 3-D geometries of Micromegas have been built with a few simplifications. For example, inter penetrating wires are taken to build the calendared mesh. A weaving model is however built for a woven micro-mesh with the help of the geometry tools available in COMSOL. Moreover, an asymmetric woven mesh (where the X wire and the Y wire does not have the same amplitude of weaving) can also be built from the same model by tuning the parameters. Since the companies sometimes provide such asymmetric woven meshes, modeling their geometry is also an interesting study. 
In this entire study, the anode, the mesh and the cathode are built on XY planes. Generally, the center of the mesh plane is placed at Z=0, the anode is placed in the negative Z direction and the cathode is placed in positive Z direction.  The same convention will be followed in all the line-plots and surface-plots. The amplification gap is defined from the bottom most point of the mesh till the anode. Similarly, the drift gap is defined from the top most part of the mesh till the cathode. This definition is specially important for the woven meshes.

For both, field calculation and discharge modeling, we have started the problem with a simple geometry, i.e., a parallel plate. The plates are separated by the same distance as the micro-mesh and the anode plane (typically 128 $\mu$m). Solving the problem with a simple geometry is a good way to understand the results and also to disentangle the effects of sharp or narrow or curved geometries. 

\begin{figure}[h]
\begin{minipage}{14pc}
\includegraphics[width=16pc]{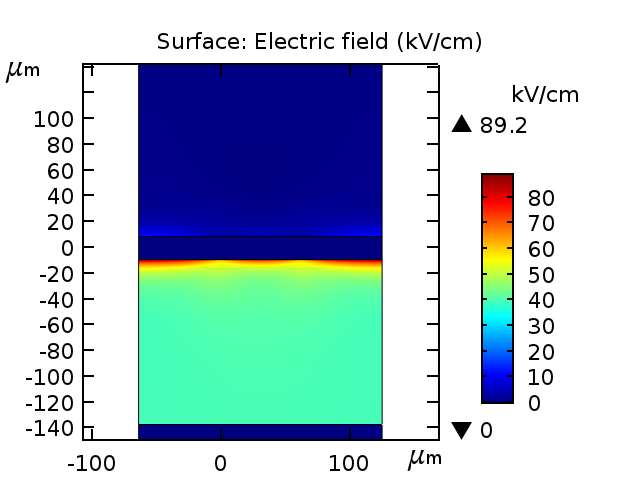}
\caption{\label{label} Surface plot of the field for 18-45 calendared mesh.}
\end{minipage}\hspace{4pc}%
\begin{minipage}{14pc}
\includegraphics[width=16pc]{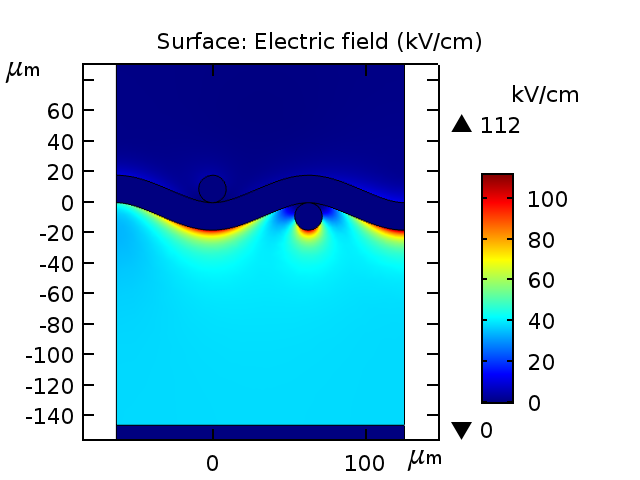}
\caption{\label{label} Surface plot of the field for 18-45 woven mesh.}
\end{minipage} 
\end{figure}

\section{Results}
 	\subsection{Electric Field}
 		\subsubsection{Parallel Plate geometry:}
 		Three parallel metal plates, each of 20 $\mu$m thickness are taken. The bottom plate, which is anode, is kept at ground potential. Another metal plate (say the middle electrode, analogous to micro-mesh), which is at -540 V, sits 128 $\mu$m above the anode. The cathode is placed 1 mm above the middle electrode and biased with -600 V. The geometry is illustrated in figure 1 (however the figure does not show the cathode). All the electrodes are in XY plane, i.e., the Z axis is going vertically through all the planes. The electric field is solved for this geometry and the field is found to be absolutely uniform. In figure 4, the surface plot of the field is made on a plane which slices all the electrodes vertically in two equal halves. The field inside the metal is zero and the field between the metal planes is very uniform. The electric filed is also plotted along the Z axis, which is going through the center of all the electrodes. Starting from the left side of the plot (figure 5), the field (analogous to drift field) has a constant value till the line passes through the middle electrode where the field goes to zero. Then in the narrow gap, the field becomes very high (analogous to amplification field) and again goes to zero as the line ends to the anode. The solved field for these ideal parallel plates is actually very identical to the analytical values. 
\begin{figure}[h]
\begin{minipage}{14pc}
\includegraphics[width=12pc]{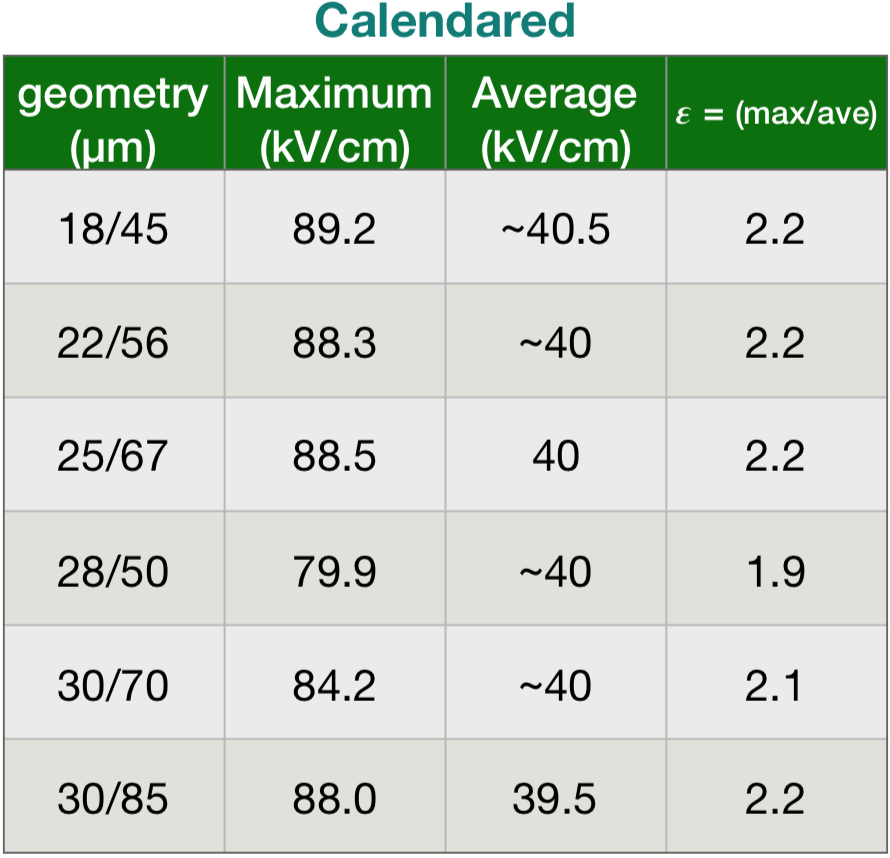}
\caption{\label{label} The maximum and average field values in calendared geometry.Mesh is at -540 V.}
\end{minipage}\hspace{10pc}%
\begin{minipage}{14pc}
\includegraphics[width=12pc]{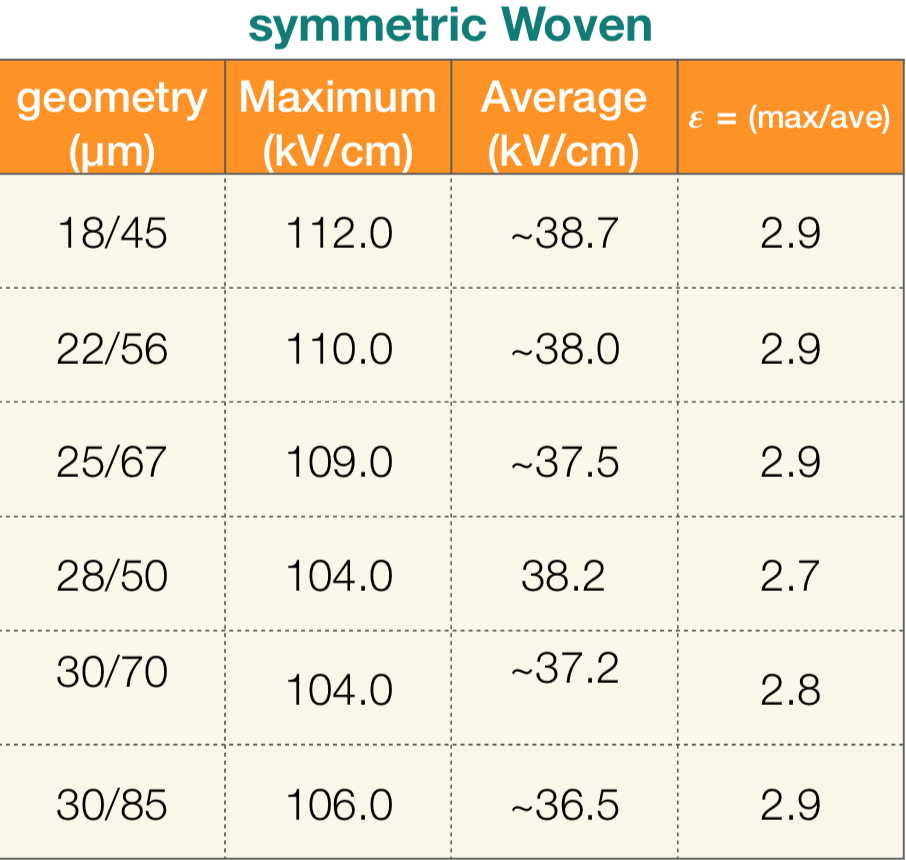}
\caption{\label{label} The maximum and average field values in woven geometry. Mesh is at -540 V.}
\end{minipage} 
\end{figure}
 		\subsubsection{Micromegas geometry:}
 		Now coming to the Micromegas geometry, two different kinds of meshes are modeled: calendared meshes and woven meshes. Inter-penetrating mesh wires are taken to build the calendared structure (figure 2). For the woven mesh (figure 3), an ideal weaving geometry is built by using the parameter based geometry tool. The amplitudes and the repetitions of weaving are first measured with a microscope. The anode is 128 $\mu$m below the mesh and the cathode is 1 mm above the mesh. The anode is at ground potential, the mesh is at -540 V and the cathode is at -600 V. 
 		To inspect the field, surface field plots are drawn (figure 6-7). For the surface field plots, the probing plane is taken perpendicular to the anode, mesh and cathode. It passes through one of the wires so that the field immediately below the wire can also be checked.
 		From the surface plots it can be seen that the field just above and below the mesh is not as uniform as we have seen in the case of parallel plates (figure 4). Compared to a flat metal surface, the field near a curved cylindrical surface becomes higher. The maximum field in the amplification region goes quite high (89 kV/cm) than the expected average field. For woven meshes, the field maximum is even higher (112 kV/cm) owing to the weaving pattern of the wires (figure 6-7). 
 		A systematic field study is performed on different calendared and woven structures. The maximum field inside the amplification gap and the average field are tabulated in figure 8 and figure 9 respectively. 

\begin{figure}[h]
\begin{minipage}{16pc}
\includegraphics[width=17pc]{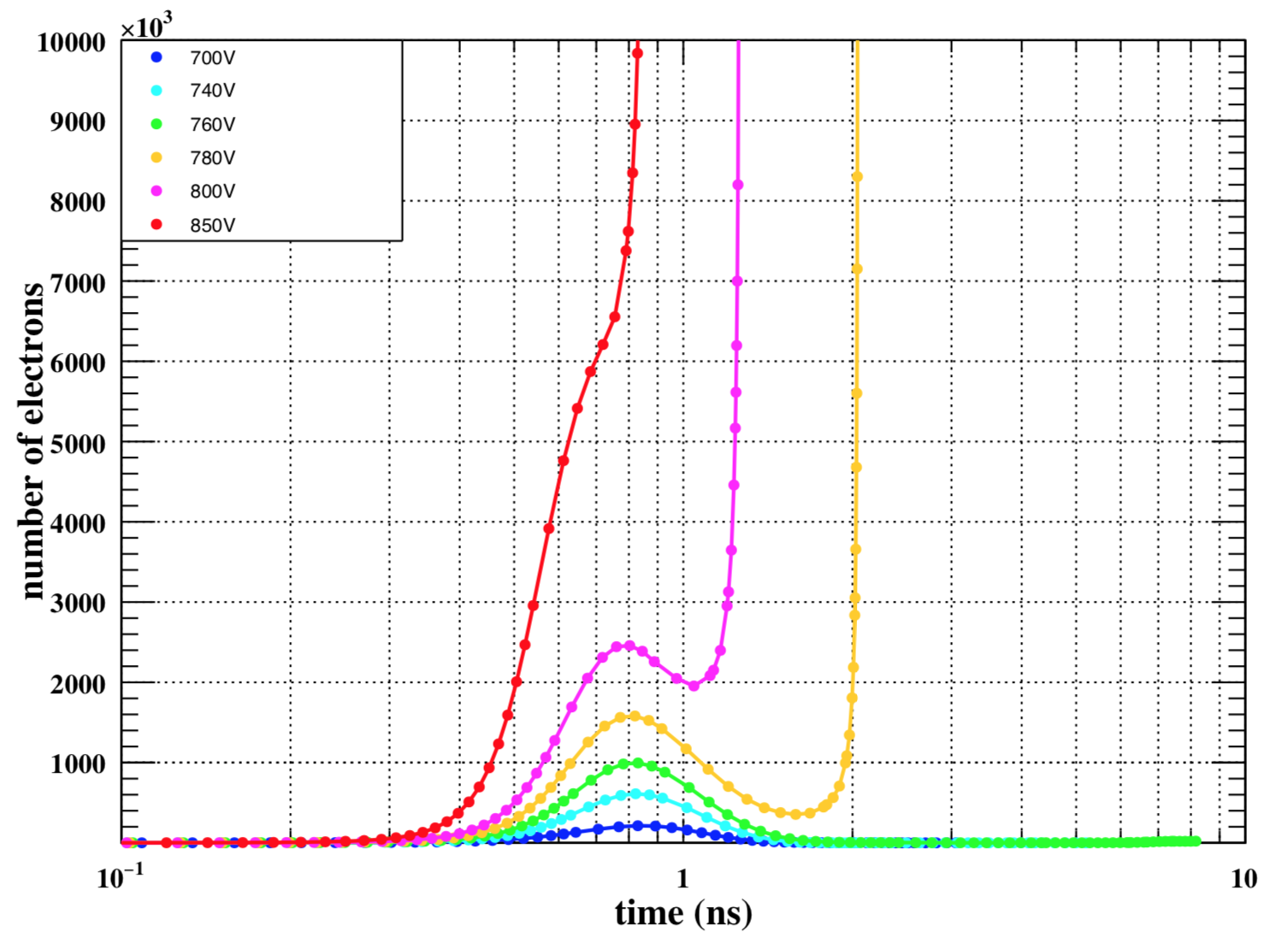}
\caption{\label{label} Development of discharge in parallel plates. A voltage scan in simulation.}
\end{minipage}\hspace{6pc}%
\begin{minipage}{16pc}
\includegraphics[width=11pc]{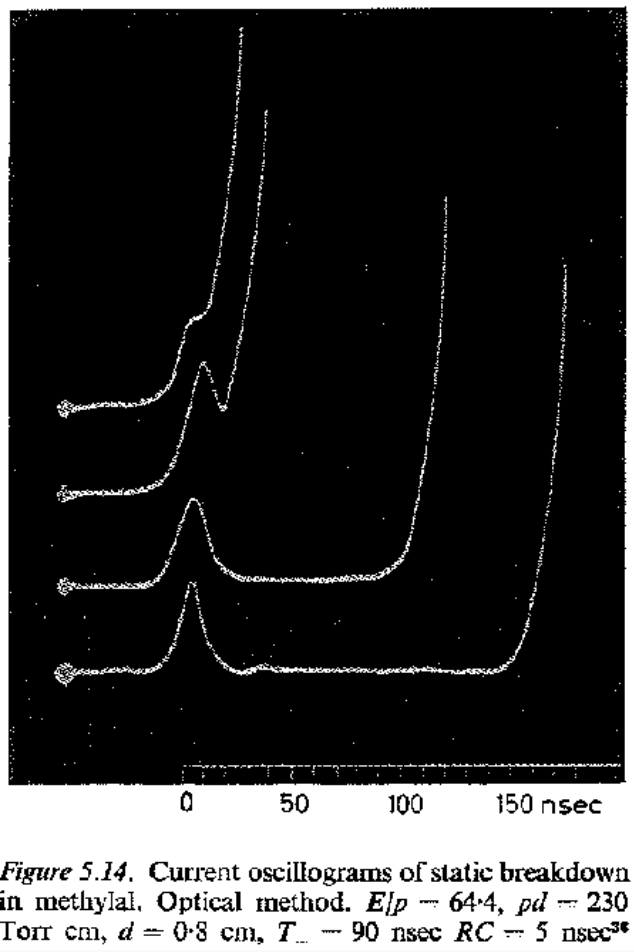}
\caption{\label{label} Experimental observation of the development of discharge in parallel plates at different voltages [12].}
\end{minipage} 
\end{figure}

 	\subsection{Discharge}
		\subsubsection{In parallel plates:}
		To study the occurrence and the development of discharge we have again started with a simpler geometry, i.e., only two parallel plates. The plates are of zero thickness and separated by a distance of 128 $\mu$m. The upper plate (cathode) is at a negative potential and the lower plate (anode) is at ground potential. The potential of cathode is varied to perform a voltage scan on how the total number of electrons changes with time for different fields. A gas mixture of 70$\%$ Ar and 30$\%$ CO$_2$ is taken. 
		Equal amount of electrons and ions are added to the volume as initial condition. The clusters have Gaussian distribution and the mean position is set 10 $\mu$m below the cathode. In figure 10, the volume integral of the electron density or the total number of electrons is plotted as a function of time. In this plot different curves represent different cathode voltages or different amplification fields. All these curves starts from the same primary cluster which contains 100 electrons. 
		The voltage scan is done from -700 V to -850 V. At -700 V, as the primary electron cluster moves towards the anode, charge multiplication starts and the total charge increases with time. When the electron cluster reaches anode, they are drained out from the volume because a $\it{out}$-$\it{flow}$ type boundary condition is applied to the anode surface. The same boundary condition is applied to the cathode surface to collect the ions. It is relevant to mention here that for this gas mixture, a penning transfer rate of 0.5 is used to calculate the Townsend coefficient with Magboltz. As the cathode voltage is increased from -700 V to -740 V and then to -760 V, the maximum of the electron-number (the maximum of the curves in figure 10) also increases. For all the three curves, eventually the electron number goes down in time as they are absorbed on the anode.

\begin{figure}[h]
\begin{minipage}{16pc}
\includegraphics[width=16pc]{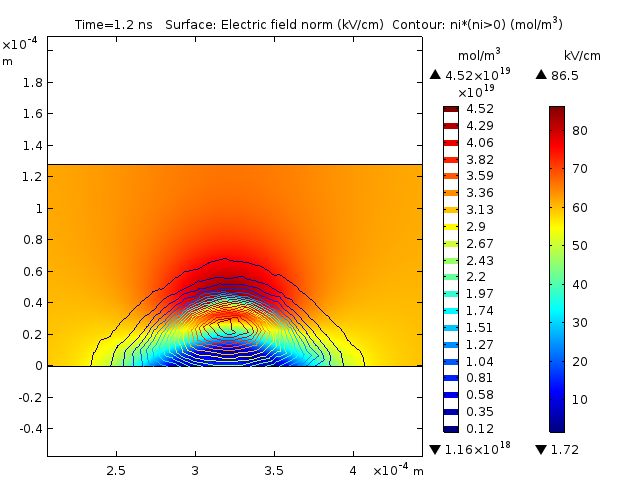}
\caption{\label{label} Change of amplification field due to space charge accumulation.}
\end{minipage}\hspace{3pc}%
\begin{minipage}{18pc}
\includegraphics[width=16pc]{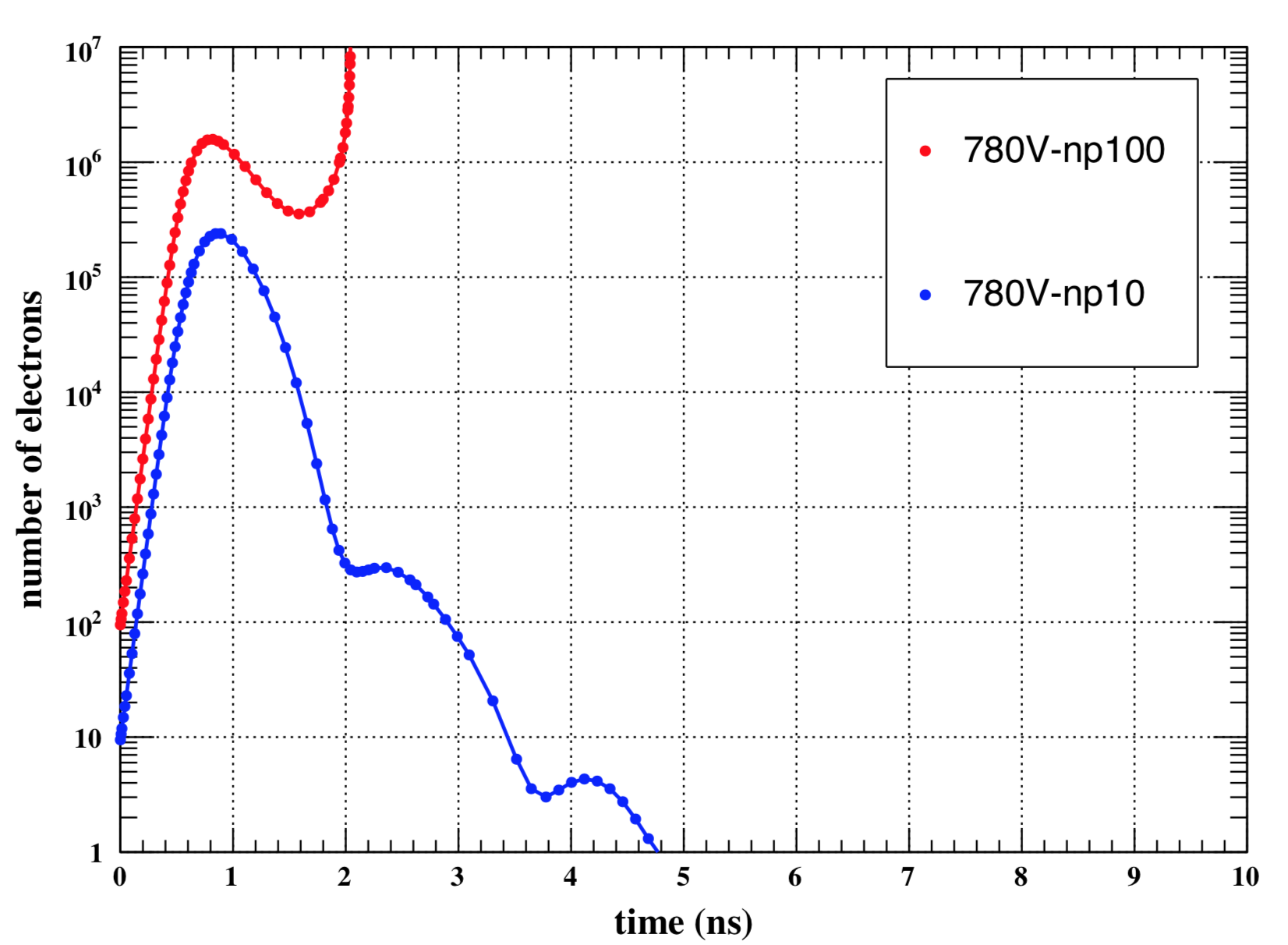}
\caption{\label{label} Development of discharge for different number of primary electrons.}
\end{minipage} 
\end{figure}

		Now, as the voltage is further increased to -780 V, the nature of the curve starts to change. With the increase of voltage, the electron and ion production increases. The electrons are fast and quickly drift towards the anode. When the leading edge of the electron cluster is mostly absorbed on the anode, the ions are far from the cathode. Therefore, the ionic space charge start to dominate in the amplification gap. This space charge obviously changes the applied field in the gap and gives rise to field distortion. The field distortion inside the amplification can be understood from figure 12. The field between the space charge and the anode is screened, where as, the field between the space charge and the cathode has increased. As the space charge moves towards the cathode, the distorted field becomes stronger in that small region. If the space charge accumulation is high enough, the distorted field becomes so strong that it can initiate avalanche. Hence the upward streamer starts. Owing to this phenomena, the end part of the -780 V curve goes abruptly high with time. If the voltage is increased further, at some voltage (-850 V for example in figure 10), the detector sparks readily. The simulation results shown in figure 10 closely match with the experimental observations on the development of discharge in parallel plates (figure 11) [12].

		\subsubsection{With different primary clusters:}
		Another study has been carried out to compare the effect of different number of primary electrons. If the calculation is started with 100 primary electrons, it is found that the parallel plate shows a streamer based discharge at -780 V.  Now the same calculation is repeated with 10 primary electrons. The results are compared in figure 13. For 10 primaries, the model does not show discharge. This can be understood as an effect of the space charge. If the calculation starts with 100 primaries, the total number of produced ions is one order higher than the case of 10 primaries. So, the effect of space charge and hence the field distortion is also higher for 100 primaries and this eventually leads to discharge. 

		\subsubsection{Effect of humidity:}
		The effect of humidity on the development of discharge has also been studied. For this study, 1000 ppm of water is added to the pure (or dry) Ar-CO$_2$ mixture and all the charge transport properties for the new admixture (say a humid mixture) are re-evaluated. In figure 14, the change of the total number of electrons with time for these two gas mixtures (dry and humid) is compared. The comparison is made at a potential difference of -700 V. The total number of electrons, for the dry gas, rises and then goes down with time. Where as, the humid gas shows discharge at the same potential difference of -700 V. 

\begin{figure}[h]
\begin{minipage}{18pc}
\includegraphics[width=16pc]{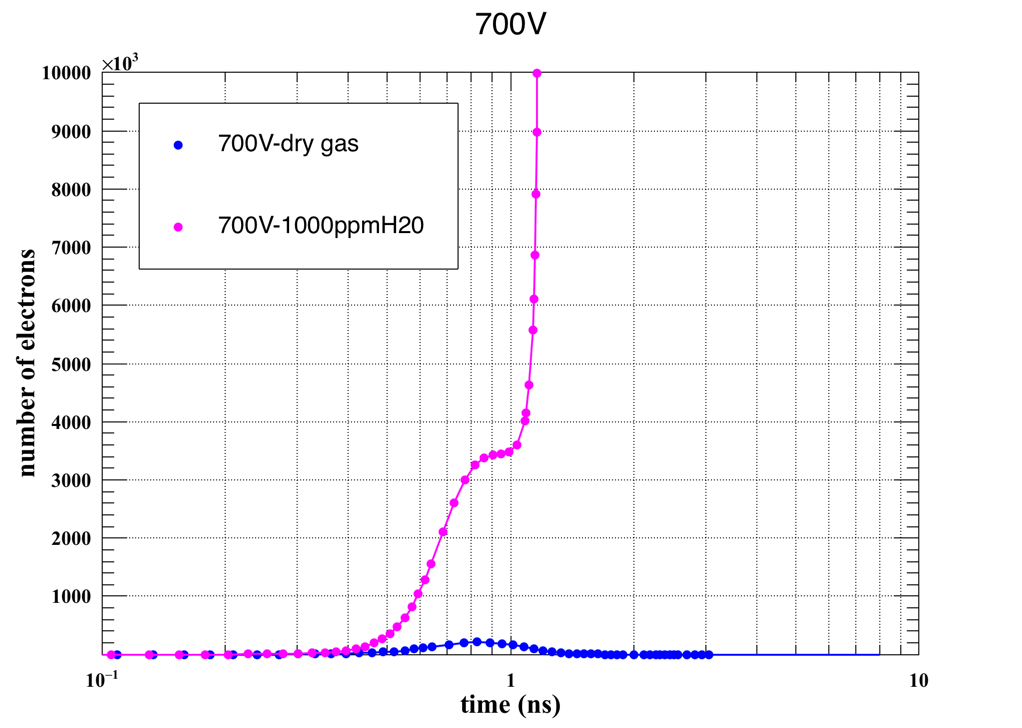}
\caption{\label{label} Change of numbers of electrons in time for the pure and the humid gas.}
\end{minipage}\hspace{4pc}%
\begin{minipage}{16pc}
\includegraphics[width=16pc]{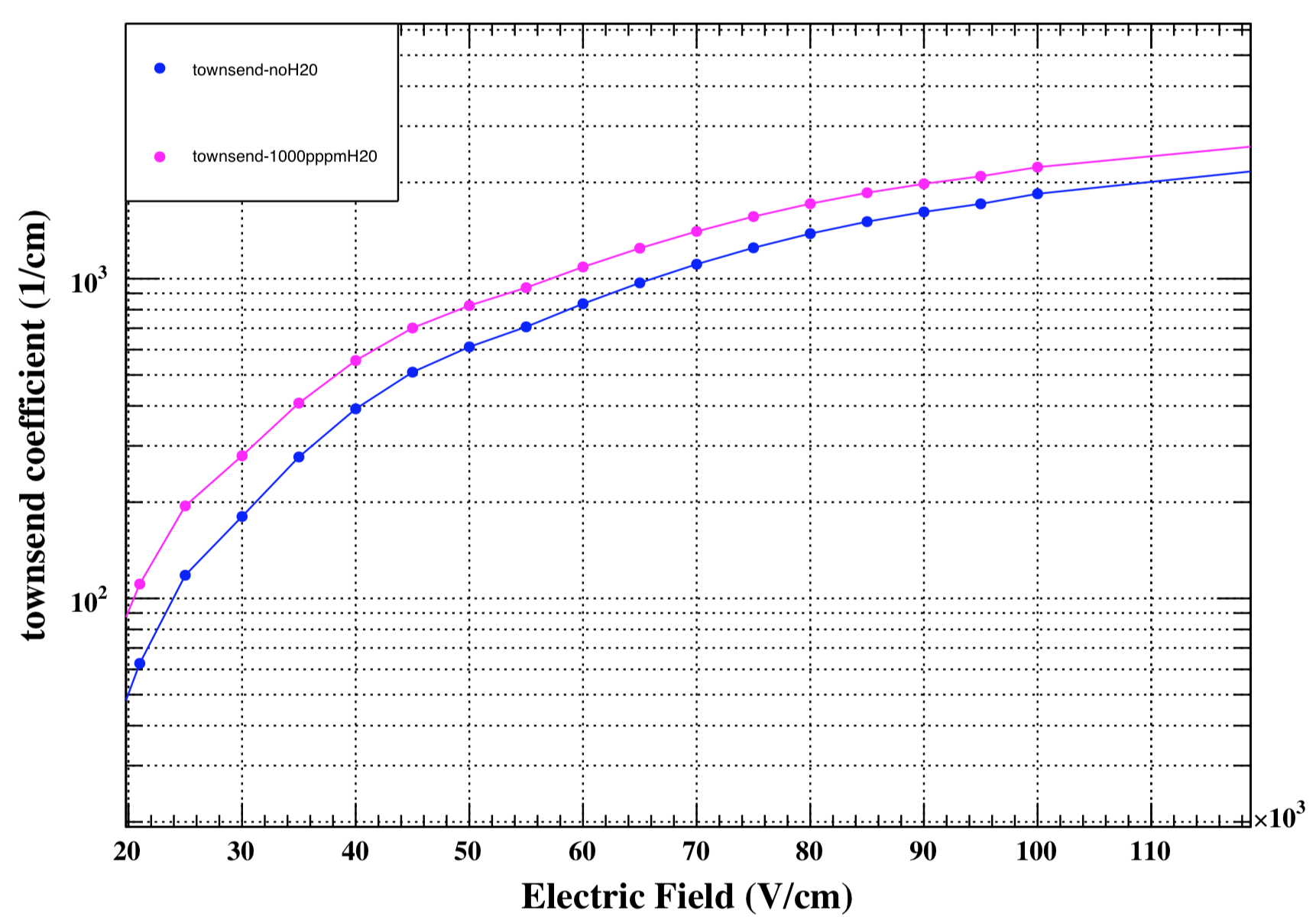}
\caption{\label{label} Comparison of Townsend coefficients for pure and humid gas.}
\end{minipage} 
\end{figure}

		The different behavior of the dry and the humid gas can be understood from their Townsend coefficients. The electric field dependence of the Townsend coefficient for both the dry and the humid gas are plotted in figure 15. At a potential difference of 700 V, i.e., for a field of 55 kV/cm, clearly the Townsend of the humid gas is higher than the dry gas. The humid gas, at 45 KV/cm, has the same Townsend as the dry gas has at 55 kV/cm. Therefore, adding 1000 ppm water translates to the reduction of field by 10 kV/cm (or reduction of voltage limit by 128 V for a 128 $\mu$m gap size). Please note, the comparison is made for a couple of ideal parallel plates where the field is very uniform.
		
		\subsubsection{In Micromegas:}
		To study discharge in Micromegas, the drift gap is taken to be 200 $\mu$m. The cathode and the anode are taken as zero thickness planes. The mean position of the primary electron (and ion) cluster is set 40 $\mu$m above the mesh-hole. By adding equal number of positive and negative charges, we also set the initial condition for the calculation. An $\it{out}$-$\it{flow}$ boundary condition is set to the cathode and the anode boundaries to drain out the ions and the electrons respectively from the volume. The same type of boundary condition is applied to the mesh boundaries for both electrons and ions. The $\it{Out}$-$\it{flow}$ condition refers to the boundaries where the diluted species can be collected. 
		A systematic scan of the amplification field is done for the 18-45 calendared Micromegas with a constant drift field of 1000 V/cm. In these studies, the anode of the Micromegas is always kept at ground potential. The threshold mesh-voltage for discharge in 18-45 calendared Micromegas is found immediately after -330 V (it sparks at -335 V). The time evolution of the electron number at -330 V is shown in figure 16. As the curve shows, at this field configuration, the electrons drift towards the mesh and nearly 40$\%$ of them are collected on the mesh. The rest pass through the mesh-hole and as they enter the amplification gap, multiplication starts. The maximum electron number reaches to 60 and then the number goes down with time as the electrons are collected at the anode. 
		Studies of the space charge development in the case of sparks show that for the mesh geometry the high field near the bottom of the mesh-wires triggers the discharges. 

\begin{figure}[h]
\begin{minipage}{14pc}
\includegraphics[width=16pc]{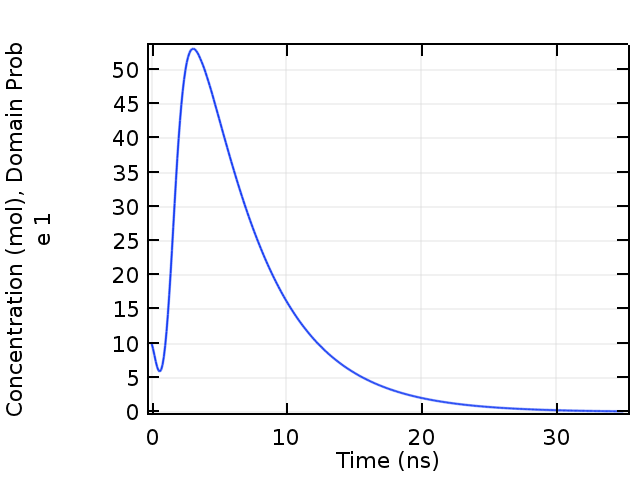}
\caption{\label{label} Change of the electrons number with time in 18-45 calendared Micromegas.}
\end{minipage}\hspace{4pc}%
\begin{minipage}{16pc}
\includegraphics[width=16pc]{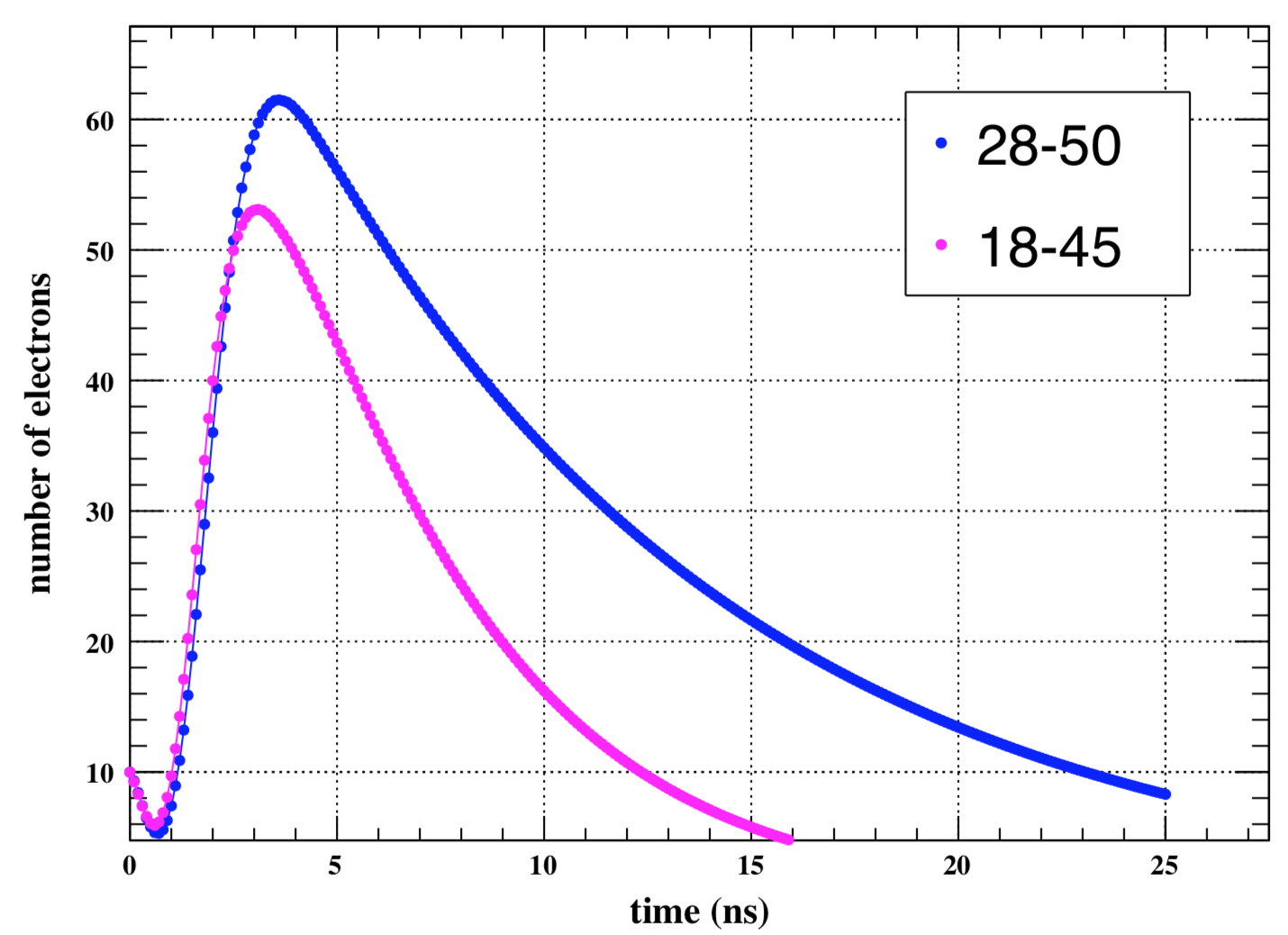}
\caption{\label{label} Comparison of the change of electron numbers with time between 18-45 and 28-50 calendared Micromegas.}
\end{minipage} 
\end{figure}

		A similar study is made on the 28-50 calendared Micromegas geometry. The change of the number of electrons with time is plotted for both 18-45 and 28-50 in figure 17 for a comparison. Both start with the same number of primaries and are operated at -330 V. The maximum number of electrons in 28-50 reaches 20$\%$ more as compared to the 18-45 geometry. Therefore, it is safe to say that 28-50 calendared behaves slightly better than 18-45 calendared. In figure 18, the transport of the electron cluster is shown at different time.

\begin{figure}[h]
\begin{minipage}[b]{20pc}
\includegraphics[width=36pc]{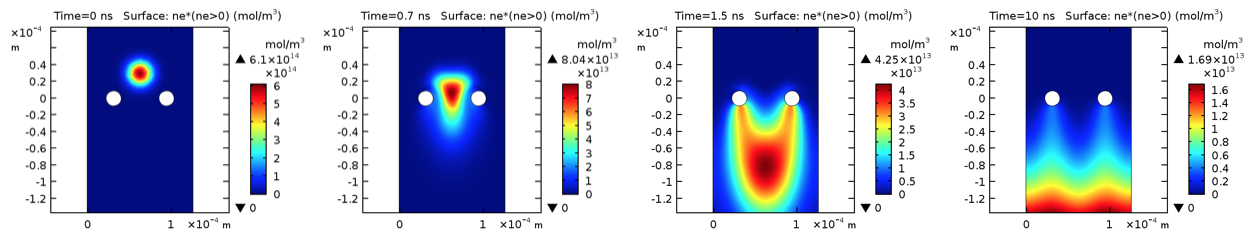}%
\caption{\label{label}transport of the electron cluster.}
\end{minipage}
\end{figure}

\section{Conclusion}
The nature of the electrostatic field in different Micromegas geometries has been studied. The maximum field, near the bottom of the wires, is quite higher than the expected average field. The maximum field is even higher in woven meshes, which suggests that calendared meshes would be a better choice in terms of HV stability. 
A simple hydrodynamic model of discharge has been developed within COMSOL + Magboltz framework. Development of discharge is studied in parallel plates. Both humidity and higher number of primary electrons enhances the discharge probability. Occurrence of discharge in 18-45 calendared and 28-50 calendared Micromegas has also been studied. 28-50 calendared behaves slightly better than 18-45 calendared.

\section{Acknowledgment}
We thank Filippo Resnati and the members of the RD51 community for their invaluable suggestions and support. We are also very thankful to BMBF for funding.

\section*{References}

\medskip

\numrefs{99}
\item 	M. Hoch, Nucl. Instrum. Method A 535 (2004) 1  
\item   CERN-LHCC-2013-006 / ATLAS-TDR-020 
\item  	CERN-LHCC-2015-012 / CMS-TDR-013 
\item  	CERN-LHCC-2013-020 / ALICE-TDR-016 
\item 	T. Behnke (ed.) et al., International Linear Collider TDR: Detectors vol. 4 (2013) 
\item 	Y. Giomataris et al., Nucl. Instrum. Meth. A 29 (1996) 376  
\item 	Comsol Multiphysics. http://www.comsol.co.in/ comsol-multiphysics. 
\item 	S.F. Biagi, Nucl. Instr. Meth. A 421 (1999) 234 
\item 	P Fonte, RD51-Note-2011-005/14-03-2011  
\item 	F. Resnati, RD-51 Open Lectures - 12/12/17 - CERN  
\item   F. Sauli, Nucl. Instr. and Meth., A386, 531 (1997). 
\item   H. Raether, “Electron avalanches and breakdown in gases,” Butterworth, London, 1964. 

\endnumrefs

\end{document}